# The increase of the light collection from scintillation strip with hole for WLS fiber using various types of fillers.

A. Simonenko[1,*], A. Artikov[1,2], V. Baranov[1], J. Budagov[1], D. Chokheli[1,3], Yu. Davydov[1], V. Glagolev[1], Yu. Kharzheev[1], V. Kolomoetz[1], A. Shalyugin[1], V. Tereschenko[1]

[1]Joint Institute for Nuclear Research, Russia, Dubna, Joliot-Curie 6, 141980
[2]NPL, Samarkand State University, Uzbekistan, Samarkand, University Boulevard 15, 703004
[3]IHEP, Tbilisi State University, Georgia, Tbilisi, University str. 9, 0186

## Abstract

The light collection of the extruded scintillator strip samples with WLS fibers placed in the longitudinal hole in the plates was measured. The holes were filled with various liquid fillers. Measurements were carried out under irradiation by cosmic muons. The method of pumping liquid filler with viscosity more than 10 Pa*s in the strip's hole with WLS fiber inside was designed and successfully tested.

---

[*]Corresponding author.

E-mail address: simonenko-av@yandex.ru, simonenko@jinr.ru (Aleksandr Simonenko)

**Introduction**

Currently, detectors based on extruded plastic scintillators are an integral part of most physical experiments in particle physics. One of their advantages is the fast rise time of the signal $\tau_{rise}$ equal to a few nanoseconds. Another important fact - this is a relatively low cost material. Such scintillators are made, usually in the form of long (several meters) plates [1, 2]. The light usually occurs via WLS fibers, which absorb the light emitted by the scintillator's material, and re-emit it in the range close to the maximum spectral sensitivity of the photo-detector. Often WLS fiber is fixed with optic adhesive on one face over the entire length of the scintillator [3].

However, the more technological solution for the placement of fibers is to provide extruded scintillators with holes passing inside the scintillator along its entire length [4, 5]. Typically, the hole diameter of 2-3 times higher than the fiber diameter. In such scintillators WLS fibers are inserted into the holes and the light from the scintillator is captured by them through an air gap.

When using a fairly long strips with WLS fibers inserted in the holes may not be sufficient amount of light entering to the photodetector. An adhesion the fibers into the inner hole may increase the light collection [4]. However, the high viscosity and limited time use of a two-component adhesive make the task of filling holes difficult. Decision in such a situation can be filling holes by suitable liquids with low viscosity or the use of optical adhesives without hardener, which eliminates the condition of time (speed) injection. In this paper we present the results of tests with different fillers. Four types of fillers were selected: distilled water, an aqueous solution of glycerol, UV glue with ultra-low viscosity "Spectrum-K-59-EN" [6] and a low molecular weight rubber "SKTN-MED" mark E [7].

Their characteristics are shown in Table 1. It should be noted that at this stage of research the fillers haven't be checked on their radiation resistance and possible chemical influence to the scintillator.

**Table 1**. Characteristics of the fillers

| Name | distilled water | aqueous solution of glycerol | UV glue «Spectrum-K-59-EN» | low molecular weight rubber "SKTN-MED" mark E |
|---|---|---|---|---|
| refractive index. (20$C^0$) | 1.333 | 1.388 | 1.460 | 1.606 |
| dynamic viscosity, mPa*c | 1 | 20 | 20 | 10000 |
| comments | | 43% solution | | hardener not used |

**Apparatus and materials**

Tests were carried out with triangular samples of scintillation strips (33mm base, height 17 mm), 50 cm long, with longitudinal holes diameter 2.6 mm, produced in the ISMA (Kharkiv, Ukraine). The strips are made by extrusion of polystyrene with additives 2% PTP and 0.03% POPOP. The sample surface was covered with a reflective layer of titanium oxide (TiO2). The sample ends were polished and covered with a layer of mirrored Mylar. WLS fiber cladding Kuraray Y11 (200) [8] 1.2 mm diameter was used, which was fixed in the hole from both ends of the scintillator with glue. Holes with thread and plastic plugs for injection of fillers on the surface (base) strips were made (Figure 1).

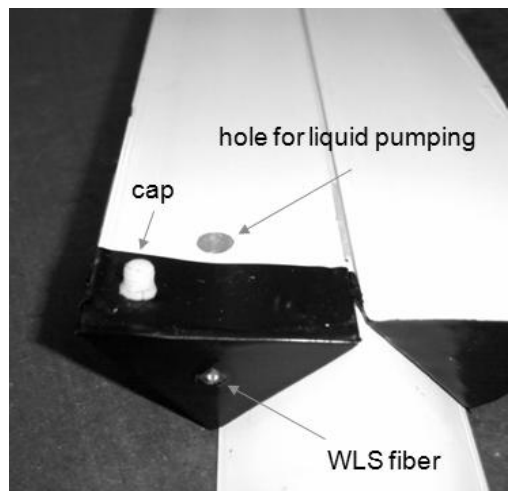

**Figure 1.** Scintillation strip samples.

We used PMT EMI 9814B as a photodetector with photocathode diameter - 51 mm. Trigger counters are based on SiPM SensL 3x3 $mm^2$ with a scintillator dimensions 20*20*20 $mm^3$ (Figure 2). These counters have the output signals in analog and digital formats.

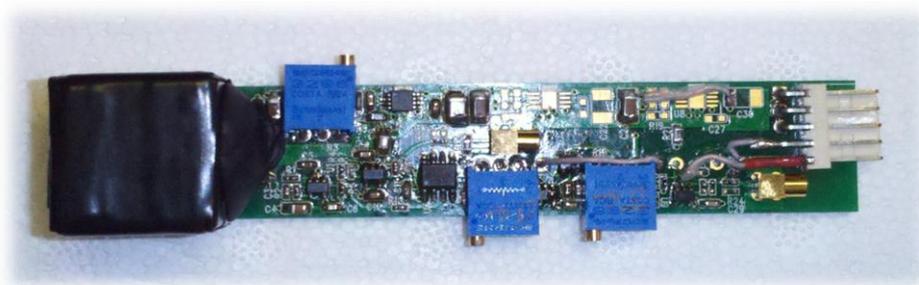

**Figure 2.** Trigger counter based on SiPM SensL.

**Test conditions**

Measurements were carried out by comparing the light collection of the same strips dry and filled with a certain filler when passing through them cosmic-ray muons. Light came only from fibers which had optical contact with PMT window through the optical lubricant. There used 2 pair of trigger counters, i.e. spectra were recruited at two points simultaneously (Figure 3). Scintillators trigger counters were adjusted in the center across the strip, thus blocking 20*20 mm$^2$ area of the test strip. Fillers pumped into the hole in two ways: liquid (water, glycerol, and the UV glue) - manually with a syringe; viscous rubber - with compressor and fluid dispenser (see below).

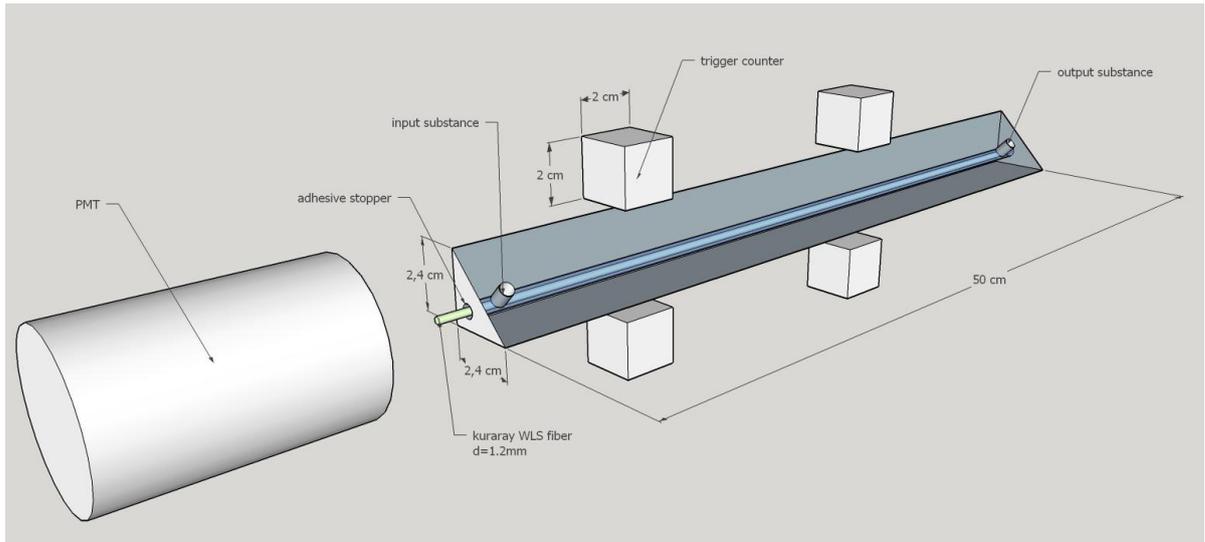

**Figure 3.** The layout of the elements.

An absolute calibration method was applied to calculate the light collection into photoelectrons [9]. Because the running distance of the muons in the triangular scintillator varies greatly (from 4 mm to 24 mm in leg size for normally incident muons), the range of output signals looks wider compared with in the rectangular strip (see. Fig.4).

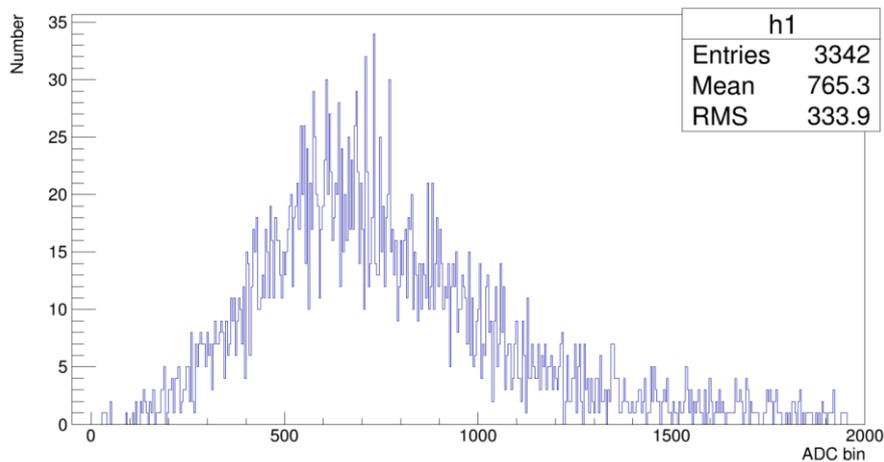

**Figure 4.** Typical spectrum of the cosmic-ray muons for a triangular strip sample.

The data acquisition system (Figure 5) was implemented in the following way. The signals from the two pairs of trigger counters after passing the discriminator, and then pairwise coincidence unit, summed and fed to the input of the gate generator. That, in turn, generate a gate signal with a specific duration (strobe) to the input of the charge-to-digital converter (LeCroy 2249W), thereby starting the processing of the signal coming from the main PMT. At the same time generated an inhibit signal for all input trigger signals. The digitized signals from ADC were read by PC, the input register showed which pair of trigger counter worked at that time moment.

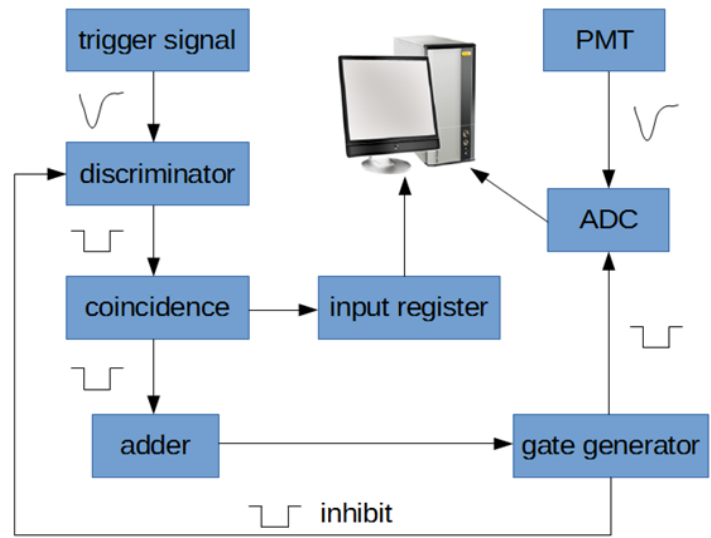

**Figure 5.**  Block-diagram of the data acquisition system.

**Methods of injection fillers**

As mentioned above, we used a conventional syringe and a transparent tube  for water, glycerin and UV glue injection to strip scintillation sample. A syringe was attached to the first hole and the tube inserted into the second hole and we squeezed until the contents began to flow out from the tube. The process continued until all air bubbles come out. At the end of the process both holes tightened by the plastic plugs.
In the case of viscous rubber used commercial compressor for supplying air to the dispenser «Fisnar» [10] and then  into a vessel with rubber. Empirically, it was picked up excess pressure in the dispenser at 0.2 atm. The rubber under the influence of such a small constant pressure slowly squeezed into a tube connected to the first hole of the scintillation strips. For filling holes strip 50 cm long it took 30 minutes. We have conducted preliminary experiments with strip samples without TiO2 layer. Results showed good filling holes with absence of air bubbles. Sealing holes after injection of rubber was carried out by adding of a small amount of hardener in both holes.

**Results**

Final results about increasing light collections for each filler are shown in Table 2. The measurements were made in four fixed positions 13, 23, 33 and 43 cm from the surface of the PMT window. For each position was got the spectrum of signals and then was defined the average value of the photoelectrons by absolute calibration. The measurement results are shown in Figure 6. The obtained data were fitted by a function $f(x) = exp^{p0+p1*x}$.

The unfilled data marked with round symbols, the data with the appropriate filler - square symbols. Priority scheduling is as follows (left - right, top - down): distilled water, an aqueous solution of glycerol, UV glue and low molecular weight rubber.

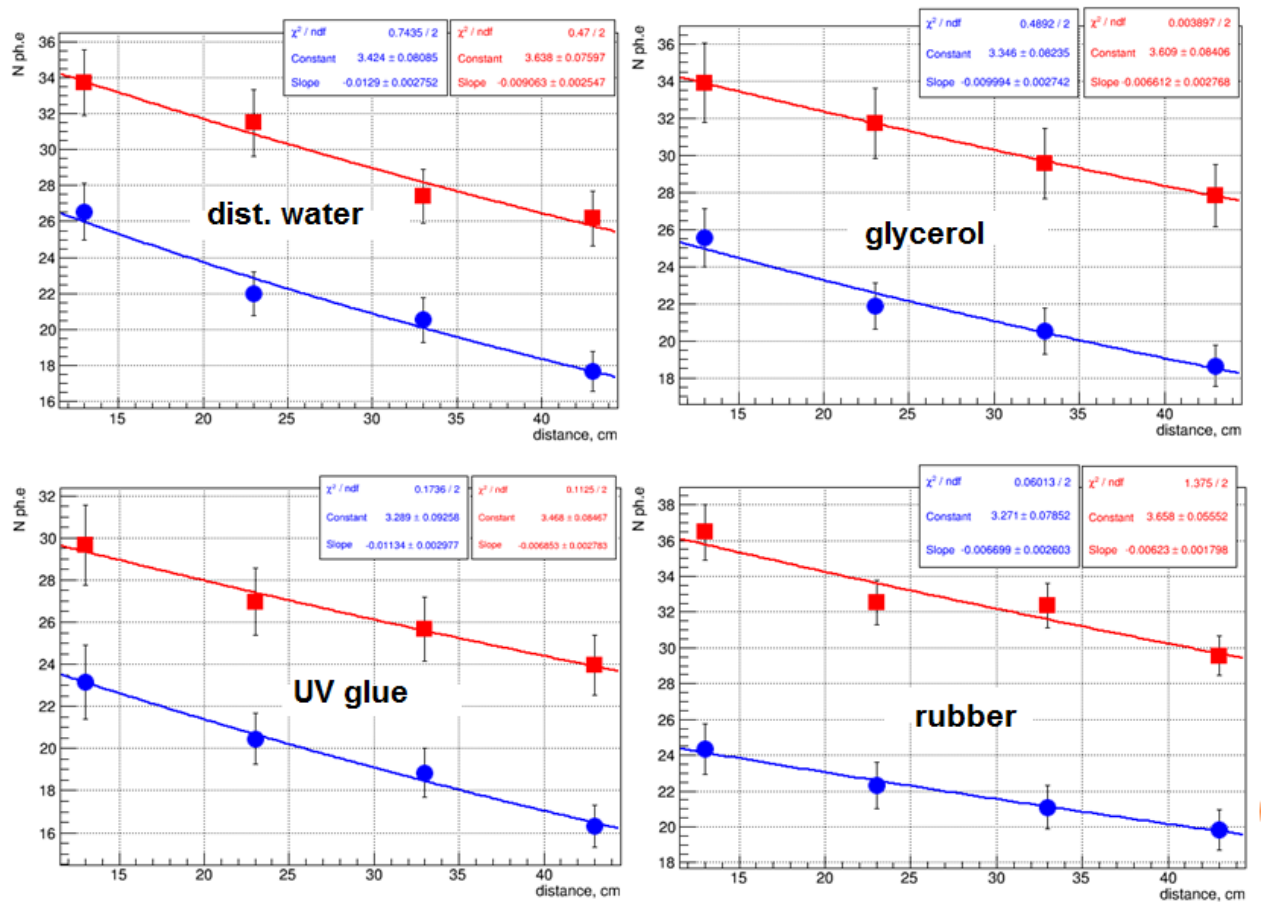

**Figure 6.** Light collection from strips for a variety of filler options. Round symbols - no filler, squares - with the filler.

**Table 2**. The total increase in light collection for each filler.

| filler | distil. water | aq. solution of glycerol | UV glue SPECTRUM | rubber SKTN |
|---|---|---|---|---|
| light collection's increase,% | 38 ± 6 | 43 ± 6 | 36 ± 6 | 50 ± 5 |

As a result, all four fluids given increase of the light collection in the range 36-50%. The greatest increase in light collection demonstrated a low molecular weight rubber "SKTN-MED" mark E (round symbols).

**Conclusions**

The samples of extruded scintillator strips with slotted holes and inserted into the WLS fibers for increasing the light collection using various optical fillers were tested.

Four sorts of the fillers have been investigated: distilled water, an aqueous solution of glycerol, UV glue ultra-low viscosity "Spectrum-to-59-EN" and a low molecular weight rubber "SKTN-MED" mark E.

The method of filler injection with viscosity greater than 10 Pa*s in the hole with a 2.6 mm diameter of the strips after placing into them the WLS fibers with a 1.2 mm diameter developed and tested. Filling time was 30 minutes for a sample strip 50 cm long.

Filling optical liquids having a low viscosity and viscous adhesives without hardener (for example in the case of rubber "SKTN-MED" mark E) in order to increase the light collection is a good alternative gluing of fibers, which is particularly problematic for the long scintillator strips.

It is shown that the use of various liquid fillers between the surface of WLS fibers and the scintillator's material enables the increase of light collection in the range of 36-50% in comparison with the samples using an air gap.